\documentclass[12pt]{spieman}  
\usepackage{amsmath,amsfonts,amssymb}
\usepackage{graphicx}
\usepackage{setspace}
\usepackage{tocloft}
\usepackage{lineno}
\usepackage{siunitx}
\usepackage{soul}
\usepackage{xcolor}
\usepackage{amsmath}
\usepackage{fancyhdr}

\newcommand{\BGal}{$\beta_{\textit{GALEX}}$}
\newcommand{\BS}{$\beta_{UVOT}$}
\newcommand{\Swift}{\textit{Swift}}

\nolinenumbers
\title{Investigating the need for a robust ultraviolet filter set aboard the Habitable Worlds Observatory\footnote{Copyright 2025 Society of Photo‑Optical Instrumentation Engineers (SPIE). One print or electronic copy may be made for personal use only. Systematic reproduction and distribution, duplication of any material in this publication for a fee or for commercial purposes, and modification of the contents of the publication are prohibited.}}

\author[a]{Kyle W. Cook}
\author[a]{Benne W. Holwerda}
\author[a]{Clayton Robertson}
\affil[a]{Department of Physics, University of Louisville, Natural Science Building 102, 40292 KY Louisville, USA}

\cftpagenumbersoff{figure}
\cftpagenumbersoff{table} 
\begin{document} 
\maketitle

\begin{abstract}

High resolution, ultraviolet imaging is often unavailable across the sky, even in heavily studied fields such as the Chandra Deep Field - South. The Habitable Worlds Observatory is one of two upcoming missions with the possibility of significant UV capabilities, and the only one early enough in development to consider suggestions to its design. In this paper, we conduct an initial study of how current common UV filter sets affect the results of spectral energy distribution fitting for the estimation of galaxy parameter. This initial look is intended to motivate the need for future, more robust, SED fitting of mock galaxies. We compare the broad near UV and far UV filters used by the \textit{GALEX} mission to the three more narrow \Swift\ UVOT filters. We find that the \textit{GALEX} filters result in larger errors when calculating the UV $\beta$ parameter compared to UVOT, and provide little constraint on the star formation age of a galaxy. We further note the ability of the UVOT filters to investigate the 2175\AA\ attenuation bump; \textit{GALEX} has a reduced capacity to trace this same feature. Ultimately, we recommend that in order to optimize the effectiveness of HWO's ultraviolet capacity for transformative astrophysics, a minimum of a FUV filter with three medium band NUV filters should be adopted. This will combine the power of \textit{GALEX}'s wavelength range with the finer sampling of UVOT around an important dust feature.

\end{abstract}


{\noindent \linkable{kyle.cook@louisville.edu} }

\begin{spacing}{2}   

\section{Introduction}
\label{s:intro}  

The spectral energy distribution (SED) of a galaxy is the composite of virtually every process occurring within that galaxy and requires high-quality ultraviolet (UV) to infrared (IR) multi-wavelength data to model appropriately. While the optical and IR regions are typically well sampled, the UV wavelengths are often from the two broad GALaxy EXplorer (\textit{GALEX}) NUV/FUV filters or are omitted completely. This is unfortunate, especially considering much of the information about the recent star formation rate (SFR) and star formation history (SFH) is contained in the UV, and drives the evolution of many other galaxy properties. Furthermore, while galaxy parameters such as stellar mass ($M_*$) are often well constrained by SED fitting, the star formation rate is typically not \cite{carnall2019, pacifici2023}. An increased availability of new, high-quality UV data will only serve to better constrain the UV end of the electromagnetic spectrum and ultimately increase the accuracy of our parameter estimation. The next opportunity to obtain UV data will come from the Ultraviolet Explorer (UVEX) mission, \cite{kulkarni2023} and it will be equipped with NUV/FUV filters similar to \textit{GALEX}. NASA's proposed Habitable Worlds Observatory (HWO) \cite{astro2020, omeara2024} provides our next opportunity to consider the utility of these filters and implement a new UV filter set that will maximize its impact on the available science. For this purpose, we conduct an initial comparison of the information that can be obtained from the \textit{GALEX} filters to the three medium band NUV filters on the \textit{Swift} Ultraviolet Optical Telescope (UVOT). In particular, we focus on the slope of the SED at UV wavelengths as parameterized by $\beta$ in the form $f_\lambda \propto \lambda^{\beta}$ \cite{calzetti1994}. This study is intended to serve as a first comparison between these filter sets to justify the need for the more computational expensive SED methods.

This paper is laid out as follows: Section \ref{s:methods} discusses the methods and data sets used preparing the data. In Section \ref{s:results} we present the results of our analysis, and in Section \ref{s:discussion} we explore the implications of these results and conclude with a recommendation on the suite of UV filters for HWO. We assume a concordance $\Lambda$CDM cosmology with $H_0$ = 70 km s$^{-1}$ Mpc$^{-1}$, $\Omega_m$ = 0.3, and $\Omega_\Lambda$ = 0.7 throughout. All magnitudes are reported as AB magnitudes.

\section{Methods}
\label{s:methods}

This study uses existing data catalogs to perform an initial comparison between the UV data from different instruments to justify further study. We examine galaxies within the well-observed Chandra Deep Field - South (CDF-S), which contains multi-wavelength photometry across the UV to IR. This field is rich with numerous existing data catalogs, and has already been identified as a target in many upcoming surveys \cite{blyth2015, ivezic2019, euclidcollaboration2024, sanderson2024}.

The first catalog used is by Zou et al. \cite{zou2022} and is a catalog of photometry and galaxy parameters from the SED fitting code \texttt{CIGALE} (Code Investigating GALaxy Emission) \cite{boquien2019}. This catalog studies three deep fields (including the CDF-S) and provides estimates for galaxy parameters such as stellar mass and star formation rate. They collect \textit{GALEX} NUV/FUV photometry from the public \textit{GALEX} database to use in their SED fitting. In this study, we use the estimated parameters as needed as well as the provided \textit{GALEX} photometry.

\begin{figure}
\begin{center}
\begin{tabular}{c}
\includegraphics[width=12cm]{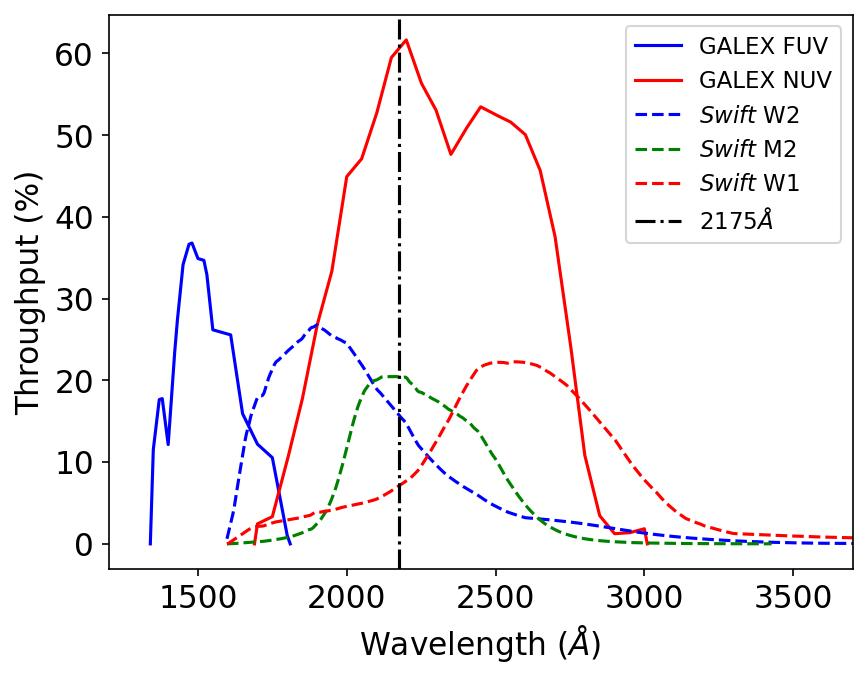}
\end{tabular}
\end{center}
\caption 
{ \label{f:filters}
The \textit{GALEX} (solid line) filter response curves and the \protect\textit{Swift} (dashed lines) UVOT filter response curves. The vertical dot-dash line is the rest frame location of the 2175\protect\AA{} absorption bump. It should be noted that this feature falls near the peak of the UVOT M2 filter.} 
\end{figure}

Nagaraj et al. \cite{nagaraj2021a, hagen2015} present an analysis of emission line galaxies and use \Swift\ UVOT photometry reduced with the default SExtractor parameters. Combining these two datasets will allow for the construction of a catalog of galaxies that have a full set of \textit{GALEX} FUV/NUV and \Swift\ UVOT W2, M2, and W1 filters. Figure \ref{f:filters} shows the response curve of each of these filters as provided by the Spanish Virtual Observatory (SVO) \cite{rodrigo2012}. Note that \textit{GALEX} offers a wider wavelength coverage, but UVOT splits the NUV filter into three more narrow filters, offering a more granular look at this wavelength range.

Both catalogs provide measured fluxes in egs/s/cm$^2$/\AA, and both are ultimately tied to the CALSPEC calibration standard. At present, PSF or aperture matching are not conducted. The corrections would likley be within the calibration errors of these missions and not strictly necessary here. When these fluxes are inevitably used in SED fitting, a more detailed analysis of the apertures will be needed.

Without both \textit{GALEX} FUV and NUV data, it is impossible to fit a value of $\beta$ for the \textit{GALEX} filters, hereafter \BGal, and sources without both filters are removed. Likewise, from the \Swift\ catalog, we remove sources that do not have a full complement of UVOT fluxes to similarly fit for \BS. We then cross-match the two catalogs to identify the same object across both data sets and reject unmatched galaxies. A final cut to the data is made so only objects with a spectroscopic redshift are retained. Spectroscopic redshifts are selected because the Zou et al. SED parameter estimations are highly dependent on redshift and this selection removes the less reliable photometric values. Later corrections conducted in this study also require redshifts and selecting the most reliable values will improve results. Galaxies with a redshift, $z_{spec} > 0.3$ are also cut to keep our sample constrained to the nearby Universe and limit issues with flux redshifting in or out of the filters studied. Additionally, the galaxies are corrected for line of sight dust extinction using the model presented by Cardelli et al. \cite{cardelli1989}:

\begin{equation}
\label{eq:cardelli}
f_{emitted}(\lambda) = f_{obs}(\lambda)10^{0.4A_{\lambda}},
\end{equation}

where $A_\lambda = a(x)A_v + b(x)E(B-V)$, $f_{emitted}$ and $f_{obs}$ are the emitted and observed fluxes respectively, E(B-V) is the color excess, and $A_v$ is the attenuation in the V-band. Following Cardelli et al., $A_\lambda$ is calculated with the following coefficients for $3.3\mu m^{-1} \leq x \leq 8\mu m^{-1}$:

\begin{equation}
    \label{eq:coefficient_a}
    \begin{split}
        a(x) &= 1.752 - 0.316x - \frac{0.104}{(x-4.67)^2 + 0.341} + F_a(x) \\
        \text{where}, F_a(x) &= 
            \begin{cases}
            -0.04473(x-5.9)^2 - 0.009779(x-5.9)^3 & \text{if } 8 \leq x \leq 5.9 \\
            0 & \text{if } x < 5.9
        \end{cases}
    \end{split}
\end{equation}

and,

\begin{equation}
    \label{eq:coefficient_b}
    \begin{split}
        b(x) &= -3.090 + 1.825x + \frac{1.206}{(x-4.62)^2 + 0.263} + F_b(x) \\
        \text{where}, F_b(x) &= 
            \begin{cases}
            0.2130(x-5.9)^2 + 0.1207(x-5.9)^3 & \text{if } 8 \leq x \leq 5.9 \\
            0 & \text{if } x < 5.9
        \end{cases}
    \end{split}
\end{equation}

where, x is inverse wavelength in $\mu m$. Values for $A_V = 0.0201$ and $E(B-V) = 0.0065 \pm 0.0004$ are from the Schlafly and Finkbiener dust maps \cite{schlafly2011} and are retrieved from the IPAC Infrared Science Archive. Errors are propagated through these calculations, and all others using the \texttt{unumpy} package. When needed, $\lambda$ is chosen as the effective wavelength of the filter in questions.

The final product is a catalog of 51 sightline corrected galaxies with a full complement of \textit{GALEX} and \Swift\ UVOT fluxes. The \Swift\ and \textit{GALEX} fluxes are each fit for an instrument specific $\beta$ in log-log space of flux vs wavelength by the following procedure. For each galaxy, a Monte Carlo resampling of the photometric errors in each filter is conducted. For each resampling, a $\beta$ is fit, and after 10,000 iterations a peak value for $\beta$ is found and the standard deviation taken as the error on the fit. This process is done for both \Swift\ and \textit{GALEX}.

The $\beta$ parameter was originally defined over the range of the International Ultraviolet Explorer (IUE) mission with a wavelength range of $1250 \leq \lambda \leq 2600$ \cite{calzetti1994}. Neither UVOT nor \textit{GALEX} cover this full wavelength range and deviations from the IUE value have been explored for each. The $\beta$ values we derived are corrected accordingly. \textit{GALEX} can be corrected to $\beta_{IUE}$ by a linear correction, however, the slope and intercept both have a redshift dependent correction \cite{battisti2016}.

\begin{equation}
    \label{eq:galex_correction}
    \begin{split}
        \beta_{IUE} &= m(z)\beta_{GALEX} + b(z)\\
        m(z) &= 1.050 - 0.395z - 2.505z^2\\
        b(z) &= -0.062 - 1.325z +10.10z^2 - 152.4z^3 + 333.9z^4
    \end{split}
\end{equation}

To correct $\beta_{UVOT}$ to $\beta_{IUE}$, one linear correction is needed to correct to $\beta_{GALEX}$, \cite{molina2020} then Equation \ref{eq:galex_correction} can be used. There is significant scattering around the $\beta_{UVOT}$ vs. $\beta_{GALEX}$ relationship, as evidenced by the the uncertainty in the slope and intercept.

\begin{equation}
    \beta_{GALEX} = (0.9 \pm 0.3)\beta_{UVOT} - (0.2 \pm 0.3)
\end{equation}

For this study, only correcting $\beta_{UVOT}$ to $\beta_{GALEX}$ would allow for a fair comparison, however we choose to correct both to $\beta_{IUE}$ as this is the standard definition of the UV $\beta$. Either option would mean that UVOT error is propagated through one additional calculation than \textit{GALEX}.

\section{Results}
\label{s:results}

\begin{figure}
\begin{center}
\begin{tabular}{c}
\includegraphics[width=12cm]{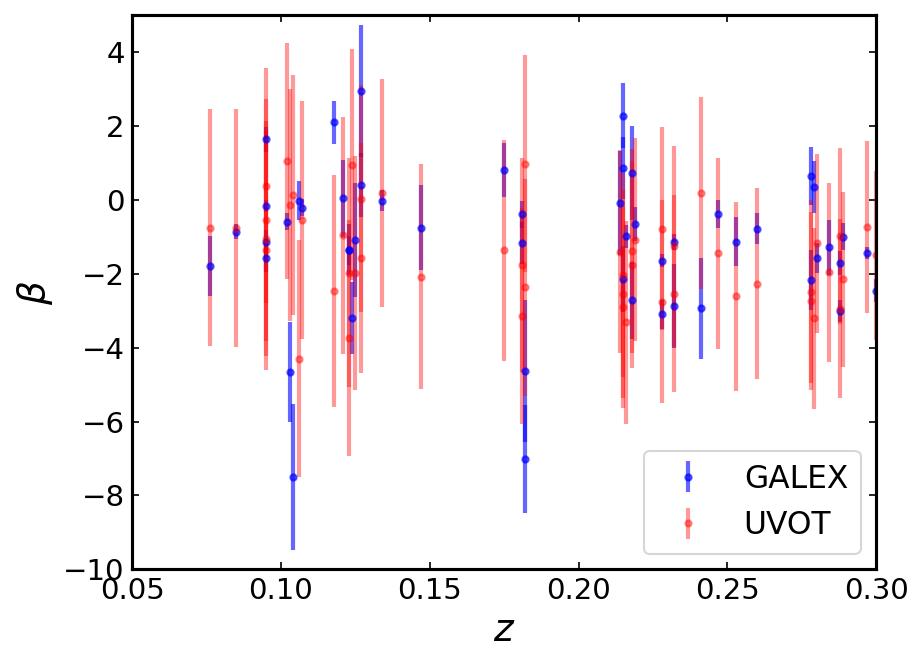}
\end{tabular}
\end{center}
\caption 
{ \label{f:betas}
The \protect\BGal\ (blue) and \protect\BS\ (red) values as a function of redshift. In most cases, the UVOT and \textit{GALEX} values agree within the error.}
\end{figure}

We first consider whether the different $\beta$ values are comparable. Figure \ref{f:betas} shows both \BGal\ and \BS\ as a function of redshift. Most notable in this figure are the large errors on \BGal\ (to be discussed in Section \ref{s:discussion}). For most galaxies, the values of \BS\ and \BGal\ agree within the error. There are some notably blue values that are highly unlikely to be physical, yet the overwhelming majority of values fall in a reasonable range, roughly $-3 < \beta < 3$. We further note that the galaxies which do not agree within the errors are primarily outside this physical range.

\section{Discussion}
\label{s:discussion}

\begin{figure}
\begin{center}
\begin{tabular}{c}
\includegraphics[width=12cm]{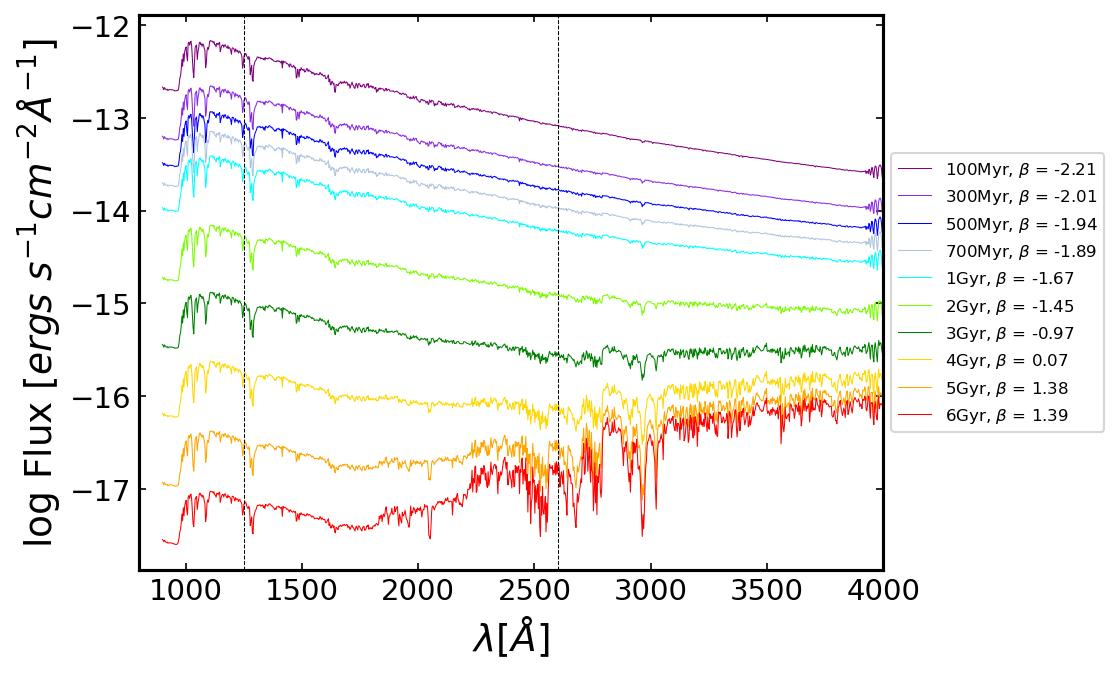}
\end{tabular}
\end{center}
\caption 
{ \label{f:models}
Models generated with B{\sc agpipes} and the corresponding values of \protect$\beta$ calculated from these models. At the top is the model with the most recent star formation "turn-on" with each model down being an older star formation age as shown in the legend. The dashed vertical lines show the range that \protect$\beta$ is fit over from the expected IUE photometry returned by B{\sc agpipes}, with the left and right line centered on the peak wavelength of the IUE bands in which \protect$\beta$ is defined.} 
\end{figure}

As previously noted, the errors on the \BGal\ values are much greater than those on \BS. We suspect two reasons for the larger errors. First, \textit{GALEX} resolution is $\sim$\ang{;;3}-\ang{;;6} providing a wider and less clean point spread function (PSF). Conversely, the \textit{Swift} UVOT resolution is $\sim$\ang{;;2.5}, offering improved PSF. Secondly, \textit{GALEX} provides two points for the fitting routine to consider, each with larger errors than UVOT. This results in a significantly larger error on the final fit. Comparatively, UVOT provides both smaller errors on the photometry and an additional point to help constrain the fit.
To estimate how much of an effect these errors have on inferred physical parameters, we use the Bayesian Analysis of Galaxies for Physical Inference and Parameter EStimation (B{\sc agpipes}) \cite{carnall2018} code to create simple model SEDs of a variety of star formation onset ages and extract $\beta$ values from them as shown in Figure \ref{f:models}. Each model is for a $10^{10}M_\odot$ galaxy with metallicity $Z = 0.5$, at a redshift $z = 0.6$ with a SFH modeled by a delayed declining exponential (delayed-$\tau$). The age of the star formation is varied to produce the different models. The choice of SFH is made to be consistent with that used in the Zou catalog. Table \ref{tab:ranges} shows the mean value over the 51 sources fit for both $\beta$ and its error. These are used to find the average minimum and maximum $\beta$ values and compare to Figure \ref{f:models} for a rough estimate of the range of star formation age. It is important to note that these are averages of the calculated $\beta$ values themselves, and not a direct value to the model $\beta$. The purpose here is to illustrate that, even after corrected, there is factor of 2 difference between $\beta_{GALEX}$, and $\beta_{UVOT}$ and examine the errors associated with these parameters as a first attempt to study how well each instrument is able to constrain $\beta$.

The average \textit{GALEX} values are $\beta_{GALEX} = 0.11 \pm 2.45$, resulting in an range of $-2.34 < \beta < 2.56$. When compared to the model SEDs, this corresponds to an estimated star formation age of less than 100 Myr to greater than 6 Gyr, or the full range of ages considered here. For UVOT the average value is $\beta_{UVOT} = 0.22 \pm 0.59$, a range of $-0.37 < \beta < 0.81$. Compared to the models, this gives a much more constrained star formation age estimate between $\sim$4-5 Gyr.

\begin{table}
\begin{center}
\caption{$\beta$ and Age Ranges}
\label{tab:ranges}
\begin{tabular}{| c | c | c | c | c |}
    \hline
     & $<\beta>$ & $<\sigma>$  & Min Age & Max Age\\ 
     \hline
     GALEX & 0.11 & 2.45 & $<$ 100 Myr & $>$ 6 Gyr\\
     \hline
     UVOT & 0.22 & 0.59 & 4 Gyr & 5 Gyr\\   
     \hline
\end{tabular}
\end{center}
\end{table}

With an eye on the future, there will be an expected reduction of the errors by the improved resolution expected with HWO. However, we also argue medium band NUV filters, such as those on UVOT, result in better estimates of $\beta$ simply by virtue of increasing the amount of data in the UV range, especially around the 2175\AA\ attenuation bump.

Figure \ref{f:filters} includes the rest frame location of the 2175\AA\ dust absorption feature, which in the SED will appear as a decrease in flux and lead to a bluer value of $\beta$. This feature falls within the \textit{GALEX} NUV filter and near the peak of the UVOT M2 filter. Previous studies \cite{battisti2016} examine the \textit{GALEX} NUV filter for a signature of the 2175\AA\ feature and conclude that it does not significantly affect in the flux in that filter, partly due to star-forming galaxies not typically displaying this feature. However, in some cases, star-burst galaxies have been found which have a 2175\AA\ attenuation bump, and may show a dependence on environment \cite{hutton2014}.

\begin{figure}
\begin{center}
\begin{tabular}{c}
\includegraphics[width=12cm]{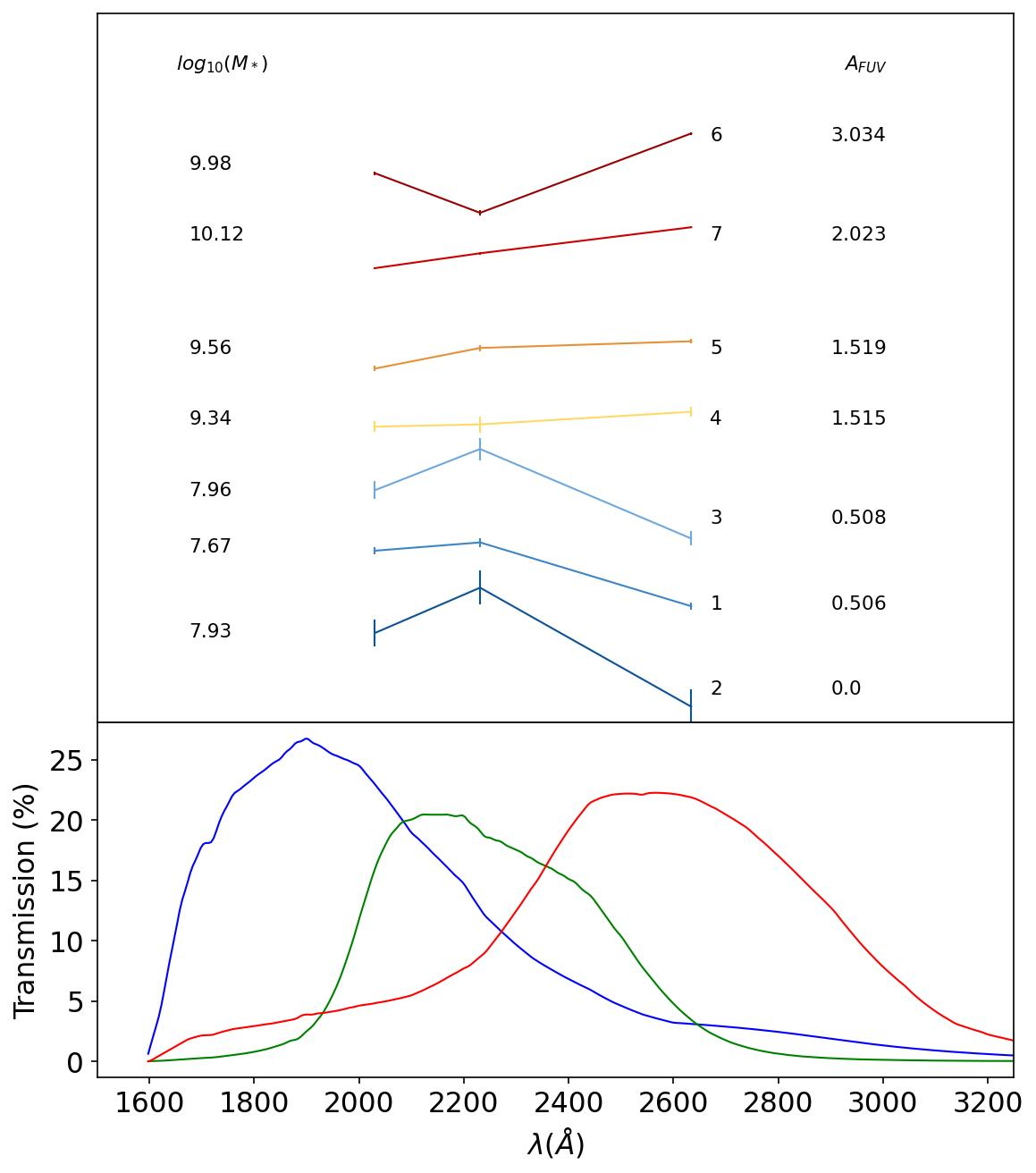}
\end{tabular}
\end{center}
\caption 
{ \label{f:colors}
Bottom: The UVOT filter response curves. Top: Seven galaxies with their magnitudes over each UVOT filter. The color of the line is a relative NUV color (W2-W1), with the galaxies being plotted with the reddest color at the top and bluest at the bottom. Additionally, stellar masses (log scale) are shown on the left and the FUV attenuation (as predicted by \protect\texttt{CIGALE} on the right). For each galaxy, the magnitudes of each filter are offset by a constant to separate the sources for easier viewing and to achieve the progression of NUV color. For this reason the y-axis in the top panel is not shown. The shape of the SED over this narrow range shows an evolution over NUV color, stellar mass, and attenuation; all are consistent with expectations for the presence or absence of 2175\protect\AA\ absorption. Numbers 1-7 are identifiers only and match the numbers in Figure \ref{f:stamps}.} 
\end{figure}

\begin{figure*}
    \centering
    \includegraphics[width=15cm]{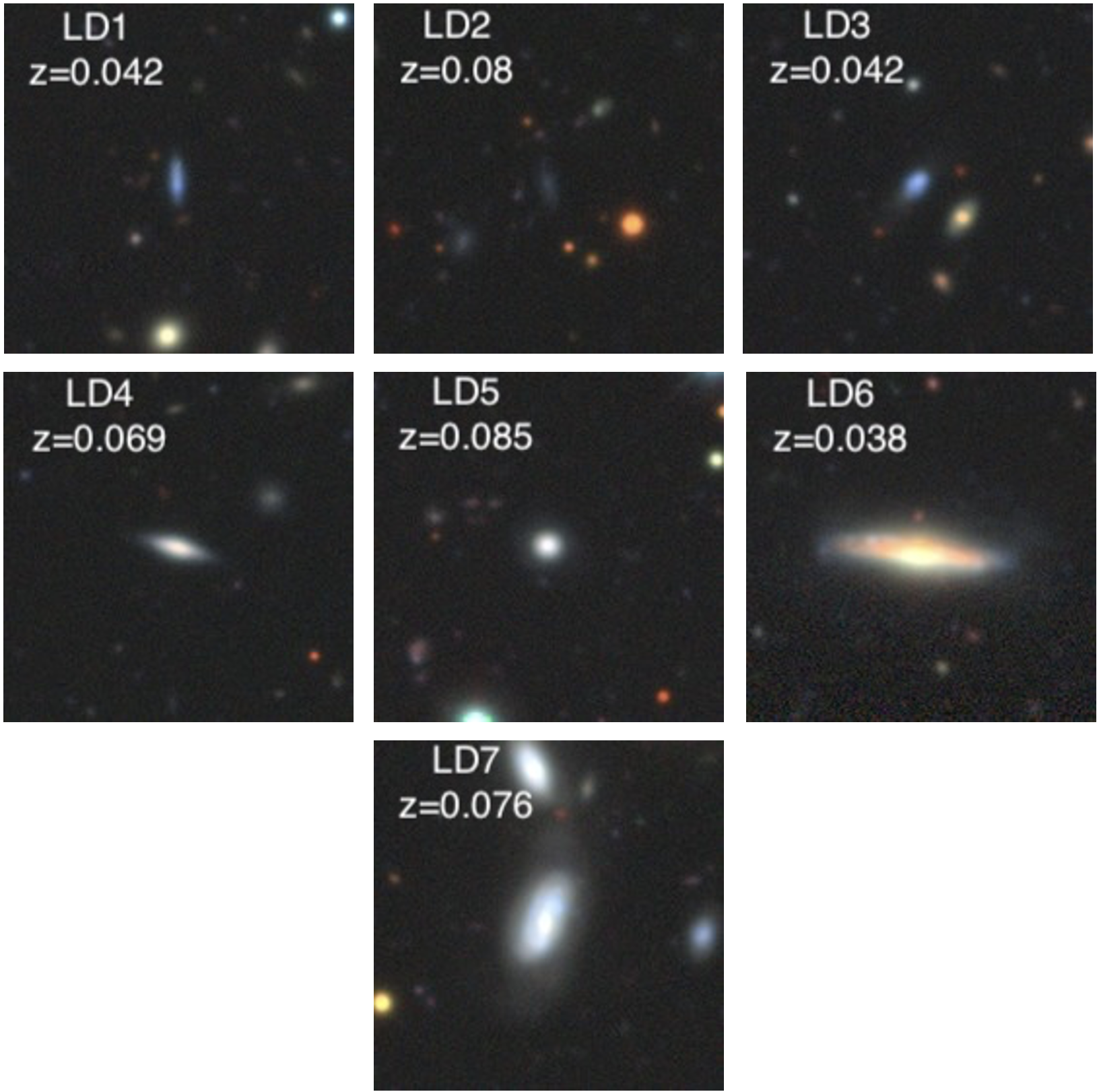}
    \caption{DESI Legacy Survey DR10 cutouts of the 7 galaxies plotted in Figure \ref{f:colors}. Numbers in the upper left corner, correspond to the numbers used previously. Redshifts are overlaid on the cutouts. Cutouts are 0.75' squares.}
    \label{f:stamps}
\end{figure*}

The effect of the 2175\AA\ feature on UVOT photometry, especially M2, has also been studied \cite{belles2023, decleir2019}. In these cases, a stronger effect on $\beta$ from the photometry is found. The tighter sampling around this feature is shown to lead to greater precision in parameter estimation. As an illustrative example, we select seven galaxies with a range of stellar masses and optical colors and compare the measured shape of the SED over the range of the UVOT filters. These galaxies are detected in neutral hydrogen by the Looking At the Distant Universe with the MeerKAT Array (LADUMA)\cite{blyth2015} survey and are additionally detected with UVOT data. The full selection will be presented in Cook et al. (in prep). Sources are numbered 1-7 in Figure \ref{f:colors} in order of increasing stellar mass. The y-axis of the top panel is not displayed because offsets are applied to the magnitudes to separate out the sources for easy viewing. The FUV attenuation, as reported by Zou et al. \cite{zou2022}, are printed on the right-hand side. The line color of each source is a relative representation of UV color (W2-W2), bluest on the bottom and increasing towards the top. The results show the bluest NUV colors and lowest stellar mass galaxies have an inverted ``V" shape, consistent with weak-to-no absorption at 2175\AA. However, as we move to redder UV color, higher stellar mass, and increased attenuation, this trend flattens and even reverses for source 6. A visual inspection of source 6 shows visible dust lanes (Figure \ref{f:stamps}); this shows an agreement between our expectation of strong absorption and a UV SED shape consistent with 2175\AA\ absorption. With only two filters, \textit{GALEX} has a reduced ability to trace this feature. Therefore, even with this simple investigation, we can see the more narrow UVOT filters can provide more information than the broad \textit{GALEX} filters, which is consistent with previous studies.

In conclusion, SED fitting routines struggle to properly recover SFRs, and the UV sky is poorly observed at present. The UV sky is dominated by recent bouts of star formation, and new high-quality UV data can improve our current modeling and knowledge of star formation in galaxies, as well as the evolutionary timescales of galaxy growth. The upcoming HWO mission is expected to be able to provide these new data products. To maximize this benefit, a more comprehensive suite of UV filters is needed instead of broad FUV and NUV bands alone. At a minimum, a FUV band coupled with three NUV bands should be included to take full advantage of the long UV coverage of \textit{GALEX} and the more precise sampling of the important 2175\AA\ feature.

We believe that \textit{GALEX}'s comparatively reduced ability to constrain the star formation age of galaxies is due both to the large errors on its photometry, and its inability to accurately sample the 2175\AA\ feature, with can effect the slope of $\beta$ when strongly present. By averaging over this feature, the broad NUV filter additionally limits our investigations of dust and PAHs in nearby galaxies. This study aims to serve as an initial exploration of the utility of these two different filter sets. With evidence in hand regarding the utility of these different filter sets, full SED sitting to mock photometry should, and will, be done to verify the initial implications found here. As of this writing, it seems that these 2 filter sets both offer important information about galaxies and combining the range and precision of them will for better recovery of galaxy parameters. The promise of HWO's power for transformative astrophysics should not be dampened by the lack of a robust filter set.

\subsection*{Disclosures}
The authors of this work declare there are no conflicts of interest, whether financial, commercial, or otherwise, that would influence the writing of this paper or the interpretations of the final results.

\subsection* {Code, Data, and Materials Availability} 
Spectral Energy Distribution results, galaxy properties, and photometry (not including \textit{Swift} are made publicly available in Zou et al.\cite{zou2022}. \textit{Swift} photometry was provided by Nagaraj et al.\cite{nagaraj2021a} via email, however, since this is not data generated by the authors of this paper, we do not make it publicly available.

\subsection* {Acknowledgments}
The work of the \texttt{astropy} community has made this work possible through the numerous packages released.

DESI Legacy Survey DR10 cutouts were retrieved using Yao-Yuan Mao's cutout tool (https://yymao.github.io/decals-image-list-tool/) and we thank him for the variety of tools he provides to the community.

This research has made use of the SVO Filter Profile Service "Carlos Rodrigo", funded by MCIN/AEI/10.13039/501100011033/ through grant PID2023-146210NB-I00.

This research has made use of the NASA/IPAC Infrared Science Archive, which is funded by the National Aeronautics and Space Administration and operated by the California Institute of Technology.

The authors further thank the NASA ADAP program for their funding of this project through grant number 23-ADAP23-0077.


\bibliographystyle{spiejour}   





\vspace{1ex}

\listoffigures

\end{spacing}
\end{document}